# How Misuse of Statistics Can Spread Misinformation: A Study of Misrepresentation of COVID-19 Data

Shailesh Bharati [1*], Rahul Batra [2]

[1*, 2] Symbiosis School of Economics,
Symbiosis International (Deemed University), Pune, India
*E-mail: shailesh.bharati@sse.ac.in

**Abstract**
This paper investigates various ways in which a pandemic such as the novel coronavirus, could be predicted using different mathematical models. It also studies the various ways in which these models could be depicted using various visualization techniques. This paper aims to present various statistical techniques suggested by the Centres for Disease Control and Prevention in order to represent the epidemiological data. The main focus of this paper is to analyse how epidemiological data or contagious diseases are theorized using any available information and later may be presented wrongly by not following the guidelines, leading to inaccurate representation and interpretations of the current scenario of the pandemic; with a special reference to the Indian Subcontinent.

## 1. Introduction

Many studies since the beginning of twentieth century (Khaleque., 2017) involving mathematical and statistical modelling of various epidemics such as Ebola, HIV/AIDS, (Zakary., 2016) and other diseases have helped estimate the probabilities of occurrences, predict the outbreaks across various geographies, provide short term and long term projection of cases in varying degrees of reliability and accuracy, and all in all have helped us understand the impact of epidemics on our lives. Policy makers have utilized these models as an important tool for decision making, where recently we saw many countries setting up lockdown in the economy as a whole or parts. Some researchers have strongly argued about the impact of social media in our lives. They have also depicted how important media can be and benefit overcome the discrepancies in passing on the information correctly to the public and help in avoiding or at least controlling such severe outbreaks (Zakary., 2016).

The data of COVID-19 outbreak is studied by several researchers and data scientists by applying different mathematical models (Rao et al., 2020; Chang et al. 2020). Many studies have been covered on this pandemic using meaningful results after applying various mathematical models. Majority of the pandemics demonstrate an exponential curve at the earlier stages of transmission and eventually flatten out (Junling et al., 2014). To study the predictions of the spread of infections due to the Novel Coronavirus, the SIR model is the best suited mathematical model. This model works on an assumption that, an individual once recovered from this infection, is unlikely to become susceptible again to the same infection (Kermack & McKendrick, 1991).

The SIR is a compartment model (Herbert, 2000) and is used to include information of individuals who may be susceptible, or infectious or recovered and deceased. Generally, an individual is counted into one of the three potential categories viz., susceptible or infected or re-covered which is denoted by their initials S, I, and R respectively. Here susceptible individual is the one who is at the risk of getting infected. The infected individuals are the susceptible individuals who get infected by the corona virus although they may be asymptomatic or symptomatic. The recovered are those individuals who have recovered, or quarantined, or died from the disease. So far, in most of the studies taken in case of India, have shown quite a significant predictive ability with respect to the growth of infections due to Covid-19 in India.

Recently, a survey revealed that ~67% of Indians trust the media for nCov-19 related news (Gulankar, n.d.). These numbers for 2019 were on average at ~44-47% with almost no changes for over two years (Grimm et al., 2016). This represents a massive change of about ~22% pts. This presents a great opportunity to mislead the general public about the ongoing pandemic, which will be divulged into later, first we shall understand how pandemics are theorized, using models; namely: SIR, SIRS, SIER, SIERS these models are explained in terms of increasing complexity which implies that SIERS model captures the current pandemic to the highest degree.

## 2. Review of Literature
### 2.1. Mathematical Modelling Methods

Many modifications have been made to the SIR and the SIS model in order to include and study other relevant variables in order to improve the model and build predictions (Anastassopoulou et al., 2020; Corman et al., 2020; Gamero et al., 2020; Huang et al., 2020; Hui et al., 2020; Rothe et al., 2020; Singh, 2020). Generally, in the SIR model, the infections from viruses which are contracted only once by individuals are accounted. Hence, susceptible individual who gets infected by the virus is subsequently removed, from the dynamics. A person once removed cannot take part again. While, in the SIS modelling, a person can be included under the susceptible category in case he gets infected again. (Tiwari, 2020) The SIQR model which stands for Susceptible, Infected, Quarantine, and Recovered is a modified version of the SIR model. This model categorizes the infected individuals into two groups, viz., those who get quarantined and others who are asymptomatic. Hence, basically in the mathematical form of the SIQR model, an extra variable "Q stands for Quarantine" is added which represent the infected individuals who show symptoms and get quarantined. Apart from SIQR, the SIRS model used the exact same equation except assuming that the immunity wanes over time (Common Cold). Another model named the SIER model adds an extra parameter 'E'. The parameter 'E' represents the exposed and represents the time difference between exposed to the disease and until symptoms show up (Infected). Another model, the SIERS model assumes waning away of immunity over time with a latent infectious disease.

Khaleque & Sen quantified the infection probability of the Ebola outbreak which happened in Liberia, Guinea and Sierra Leone, (West African countries) between 2014 and 2016, using the SIR model on the Euclidean network. They concluded that the SIR model fitted well to the data. The authors were successfully able to estimate the time / period at which the infections would peak.

Zakary et al., in their first approach, used the SIR model in order to analyse the benefits of awareness programs implemented by the government, in order to make the public aware about the dangers of the disease. They concluded that these awareness programs focused to the specified region, along with the aim to control the outbreak, were sensitive to enough contacts proportional to the high risk group. In their second approach the authors considered the inter-domain interventions of travel restrictions of the susceptible persons across high risk to low risk domains. After adding the control on travel ban of susceptible people between high risk domains and infective of high risk domains. In their conclusion, they also suggested that the social media could certainly play an important role in spreading awareness and educating the public and help in HIV/AIDS prevention.

Malavikaa et al., was able to predict the short term scenario in case of India and the States with high incidence, with great accuracy, using logistic modelling. Moreover, the SIR model was used by the authors to forecast the active cases and the period at which the curve would reach the peak and eventually flatten out. The authors suggested this model to be used for planning and preparing the healthcare system. The Time Interrupted Regression modelling was used for analyzing interventions of lockdown and their impact on the spread of active cases. The study did not find any evidence which would prove the positive correlation between the impact of lockdown and the reduction in new cases by breaking the chain of infection among the public.

### 2.2. Statistics and Graphs

CDC "Centres for Disease Control and Prevention" mentions best practices / guidelines for describing the Epidemiologic data, called as descriptive epidemiology. CDC gives very much importance to the interpretation of the data and has set guidelines for the graphical representation of the data. Some relevant ones are mentioned below. Such as the aspect ratio of graphs, appropriate scaling, representing dependent and independent variable, usage of types of lines, comparing two lines on a graph and also some relevant points to adhere to the principles of mathematics while plotting data and representation of equal units of the transformed data such as logarithmic, ranked and normalized with equal distances on the axis.

The CDC manual also mentions the guidelines for plotting propagated epidemic curves. COVID–19 falling in to the category of a propagated epidemic, it seems reasonable to follow the guidelines set by CDC so as to avoid wrong interpretations. Propagated epidemics show propagated patters and show four characteristics by the CDC (Fontaine, 2018).

**"Characteristics of Propagated Epidemic Curves**
1. They encompass multiple generation periods for the agent.
2. They begin with a single or limited number of cases and increase with a gradually increasing upslope.
3. Often, a periodicity equivalent to the generation period for the agent might be obvious during the initial stages of the outbreak.
4. After the outbreak peaks, the exhaustion of susceptible hosts usually results in a rapid downslope."

**Guidelines for representing disease rates against time.**
The CDC manual has also defined the way in which to illustrate the disease rates against time. Temporal disease rates are plotted against time taken on the x axis while the magnitude of rate of the epidemic is represented on the y axis. Usually if the rates of a disease vary more in their order of magnitude, a logarithmic scale is recommended especially for epidemiologic purposes. The CDC also suggests to the usage of logarithmic scale when researchers want to compare two or more population groups.

The CDC also specifies the guidelines to measure disease or any other health conditions on a continuous scale and no by counting directly. Although, an individuals' measure may vary over and above a specified cut off value. Hence, the CDC quotes, "To calculate incidence, special care therefore is needed to avoid counting the same person every time a fluctuation occurs above or below the cut-off point. A more precise approach involves computing the average and dispersion of the individual measurements. These can then be compared among groups, against expected values, or against target values. The averages and dispersions can be displayed in a table or visualized in a box-and-whisker plot that indicates the median, mean, interquartile range, and outliers"

Ehrenberg, A.S.C. in his work "Rudiments of Numeracy" in 1977, has summarized, "Many tables of data are badly presented. It is as if their producers either did not know what the data were saying or were not letting on." He has also talked about the difference in graphs and tables in his sixth rule studied in the same work. He claims that people usually find it easier or comfortable to read the graphs instead of tables, which is only partially true. Although, graphs fall short in representing the quantitative aspects seen in the epidemiological data, but they do communicate or make some qualitative aspects more prominent for the viewer. Qualitative aspects such as a curve rather than a straight line, if some phenomena have increased or is smaller than a larger value and so on; are represented distinguishably using the graphs. Such a claim shouldn't misled us, into thinking that one can only draw out or retain very trivial information from using a graph. In his work, he has quoted that, "Most graphs do not show simple patterns or dominant numbers which can readily be grasped. Success in graphics seems to be judged in producer rather than consumer terms: by how much information one can get on to a graph (or how easily), rather than by how much any reader can get off again (or how easily)."

The literature review studied, clearly points towards the fact that although, researchers have been using various mathematical models since earlier days, in order to study the predictions of epidemiological data, there are guidelines on how the results of those predictions should be represented using graphs and tables for right or correct interpretations by the public. Following the same lines, this paper attempts to analyse the way in which sources depict the data in various forms and what wrong interpretations are being made out of those depictions.

3. **Methodology**

Due to unavailability of Primary Data, this Investigative Analysis uses data compiled by various sources such as ICMR (Indian Council of Medical Research, New Delhi), MoHFW (Ministry of Health and Family Welfare, Government of India), and various State welfare sources. We consulted data for the number of cases detected in the three states Maharashtra, Kerala and Karnataka, because the mathematical many researchers have performed mathematical modelling showing the predictions of the outbreak in India; for these three states, using Logistic Growth and SIR Models'. Hence, we have use the values imputed in these papers for actualising the models. The data available for India is from 30$^{th}$ January 2020 to 28$^{th}$ August 2020 (197 days) and the data available for the States of Maharashtra, Kerala and Karnataka is from 14$^{th}$ March 2020 to 28$^{th}$ August 2020 (168 days). The summary of the statistics is presented in the appendix. We have studied the available data for total number of cases as a function of time *(t)* along with other variables such as Total Confirmed Cases, Daily Confirmed, Daily Recovered, Daily Deceased, Total Recovered and Total Deceased.

We would like to express our gratitude to *the Covid-19 India Org Data Operations Group* for providing data using research friendly APIs, the Organization is appropriately cited as well. The data was then treated using Python 3.x using appropriate SciPy Packages keeping in line the objectives of the research, similarly all visualizations found in this paper are also done using the same. All visualisations are strictly generated by Matplotlib, we have also tried our best to maintain uniformity in order to compare graphs wherever we felt necessary.

## 4. Theoretical Framework and Analysis

It is important to understand that all of these models focus only on the susceptible part of the population or alternatively assume that the entire population is susceptible to the virus / disease.

### 4.1. The SIR Model

The SIR "Susceptible Infected Recovered" Model was presented by W. 0. Kermack and A. G. McKendrick (Kermack & McKendrick, 1933, p. 110) and is mathematically expressed as follows:
's' = number of individuals who are susceptible
'i' = number of individuals who are infected
'r' = number individuals who are removed (i.e., including re-covered and dead)
These can be expressed as Ordinary Differential Equations as follows:

$dS/dt = -\beta SI/N$

$$\frac{dI}{dT} = -\frac{\beta SI}{N} - \gamma I$$

$$\frac{dR}{dT} = \gamma I$$

Where; $s = \frac{S}{N}, i = \frac{I}{N}, r = \frac{R}{N}$

Substituting these in the Ordinary Differential Equations we get

$$\frac{dS}{dT} = -\beta si$$

$$\frac{dI}{dT} = -\beta si - \gamma i$$

$$\frac{dR}{dT} = \gamma i$$

Hence s + i + r = 1 is constant, we use this equation to check for faults in the simulations
- $\beta SI/N$ is the rate at which the susceptible population encounters the infected population. $\beta$ is a model parameter with units of 1/units per day
- $\gamma i$ is the rate the infected population recovers (this model assumes resistance to further infection). I is the size of the infected population
- $R_0 = \beta/\gamma$ (Basic Reproduction Rate)

For reference, we ran the parameters for Influenza as the characteristics of the disease fit the characteristics of the SIR Model, we use the parameters provided in (Mahaffy, J. M., 2018)

Graph 1: SIR model of SARS-nCOV-2019

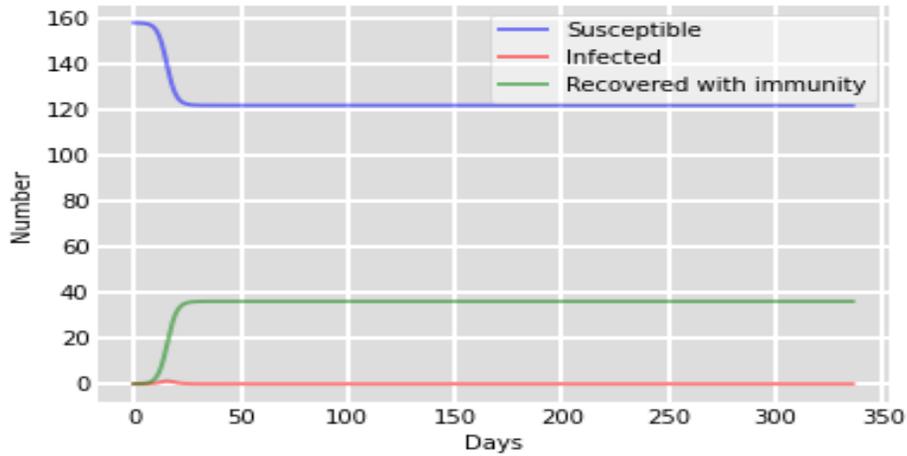

Source: *Covid-19 India Org Data Operations Group*

Although, SARS-nCOV-2019 does not fit all assumptions of the SIR model (latent period of infection) running the same simulations (Mackolil & Mahanthesh, 2020)

Graph 2: SIR model of Influenza

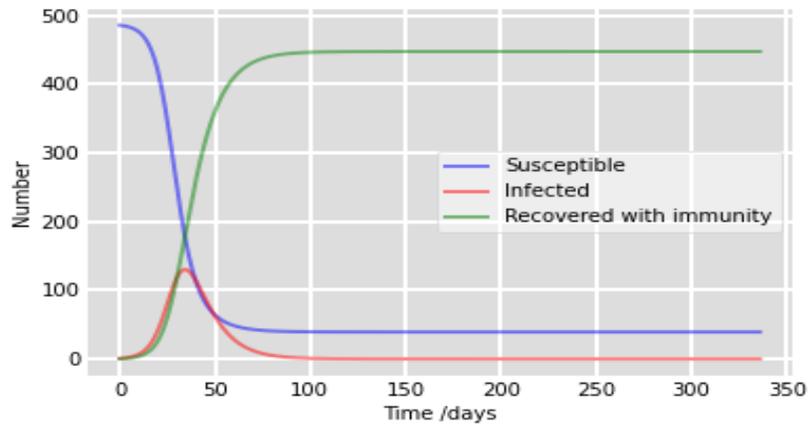

Source: *Covid-19 India Org Data Operations Group*

It seems to be extremely clear, the difference between the two graphs. The difference in the above two graphs 1 and 2, i.e., SIR Models of SARS-nCOV-2019 and Influenza is stark. When both diseases are simulated for the same period of time influenza does not even infect the entire susceptible population however SARS-nCOV-2019 infects the entire possible population. A massive difference can be seen in the $R_0$ 1.99 and 2.73 respectively.

### 4.2. The SIRS Model
This model uses the exact same equation except assumes that immunity wanes over time (Common Cold).

### 4.3. The SIER Model

The SIER model adds an extra parameter 'e'. The parameter 'e' represents exposed and represents the time difference between exposed to the disease and until symptoms show up (Infected), for SARS-nCOV-2019 it stands between 7-28 days (Lauer SA et. al, 2020)

The Ordinary Differential Equations then change to:

$$\frac{dS}{dT} = -\beta si$$

$$\frac{de}{dT} = -\beta si - \alpha e$$

$$\frac{di}{dT} = \alpha e - \gamma i$$

$$\frac{dr}{dT} = \gamma i$$

Re-running these simulations for imputed values for the State of Kerala (Mackolil & Mahanthesh, 2020)

Graph 3: SIER model for the State of Kerala

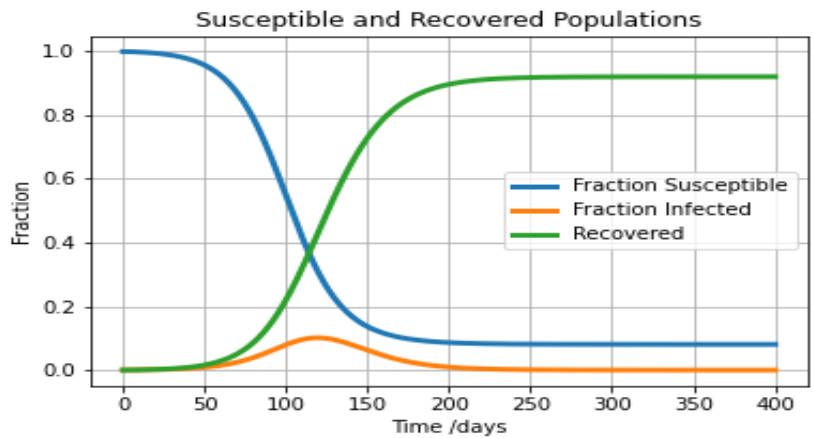

Source: *Covid-19 India Org Data Operations Group*

### 4.4. The SIERS Model
The SIERS model assumes waning away of immunity over time with a latent infectious disease.

In the next section, we introduce these models to give a reference so as to understand the graphs generated using actual pandemic data.

### 4.5. Logarithmic Scales
Before we get into the specifics of what logarithmic scale means we would first like to pose a question using graphs.

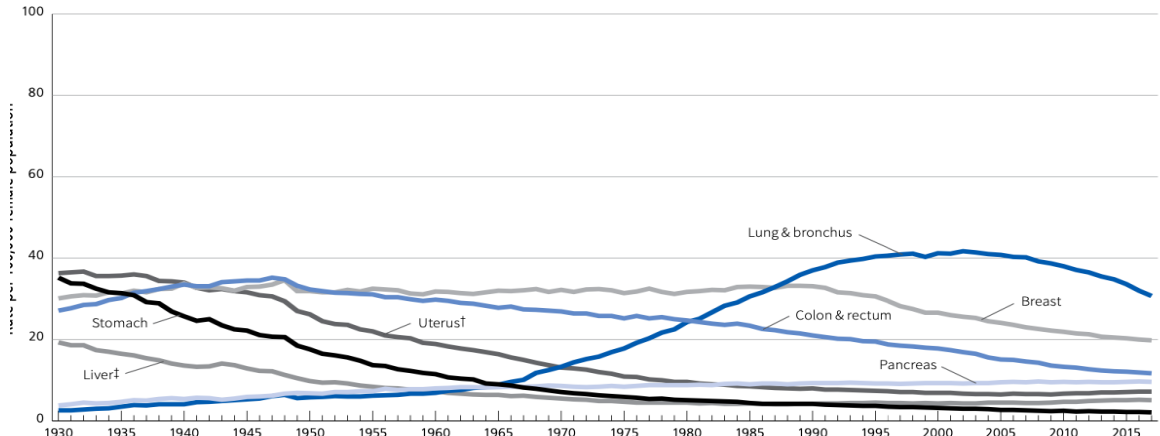

Graph 4: Types of Cancer Cases (linear scale)

Source: (Siegel et al., 2020)

In order to show the difference in type of diseases and how they are represented using different scales, in the graph above, various types of cancers are plotted across time period using the linear scale. Cancer and novel corona virus, both the diseases are extremely different in nature; cancer is not communicable in the same means as SARS-nCOV-2019. The latter represents a disease which on average spreads to 2 - 3 people (Lauer et al., 2020) and cannot be represented using the similar graphs. The CDC recommends 'semi-log' plots for the same (Fontaine, 2018)

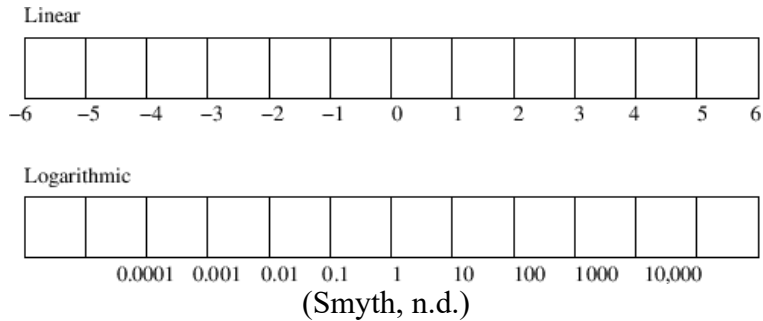

(Smyth, n.d.)

A log scale differs from a linear in terms of data represented in a unit of the graph. To make it clear, we present the outbreak of the disease in the State of Maharashtra, Kerala and Karnataka, in two different ways. The graphs on the left hand side represent the data of confirmed cases in the three states using linear scale and the ones on the right hand side represent the same data using the logarithmic scale.

Graph 5: Confirmed Cases in Maharashtra, Kerala and Karnataka (Linear and Log Scale)

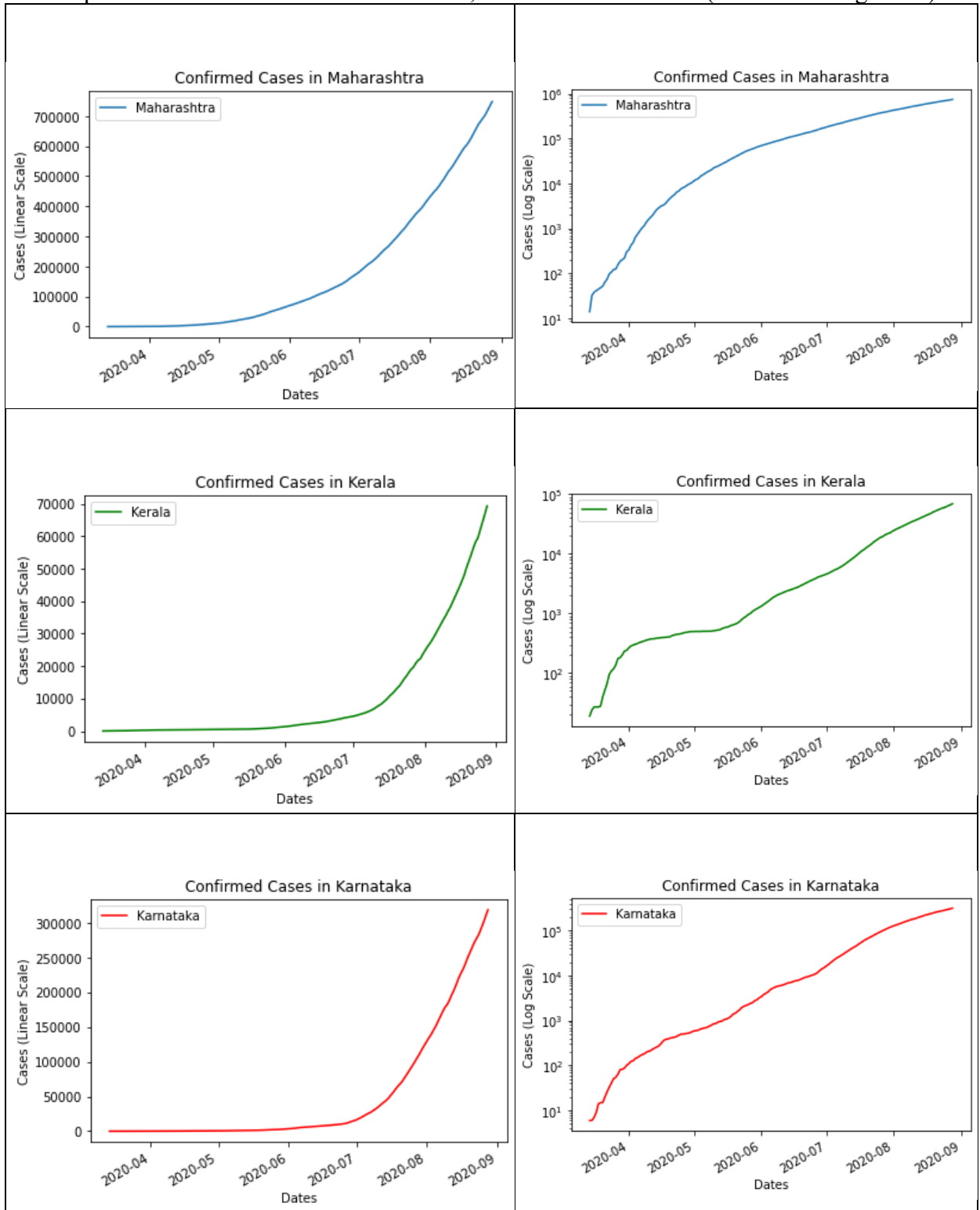

Source: *Covid-19 India Org Data Operations Group*

Same data, two different perspectives? The first graph on the left side for each states is a 'linear' graph and the second one on the right hand side is a semi-log graph, the differentiating factor being how the y-scale is depicted, the y-scale in the second graph multiplies every unit by a factor of 10 and so forth however the linear graph adds up with 1,00,000 cases per unit (Maharashtra), 10,000 cases per unit (Kerala) and 50,000 cases per unit (Karnataka). The reason we use a semi-log plot is because of the nature of these diseases, these diseases do not spread in a linear fashion and are rather exponential in nature, therefore the graphs also must take this into account when visualizing. CDC recommends the use of semi-log plot, however neither of these graphs seem to be incorrect. A reason highlighted by the London School of Economics is the fact that the public do not understand 'Semi Log' plots *however this also leads to inaccurate representation of the outbreak* (London School of Economics, 2020).

### 4.6. Not all data is created equally

Every state did not follow the same timeline i.e. First incidents of the disease were not on the same day for every state of the country. This creates an opportunity for the states who had breakouts at later stages could prepare for it beforehand. *This however, presents another opportunity for incorrect visualisation of the breakout in different states.*

Graph 6: Confirmed cases in Telangana and Arunachal Pradesh plotted against Dates

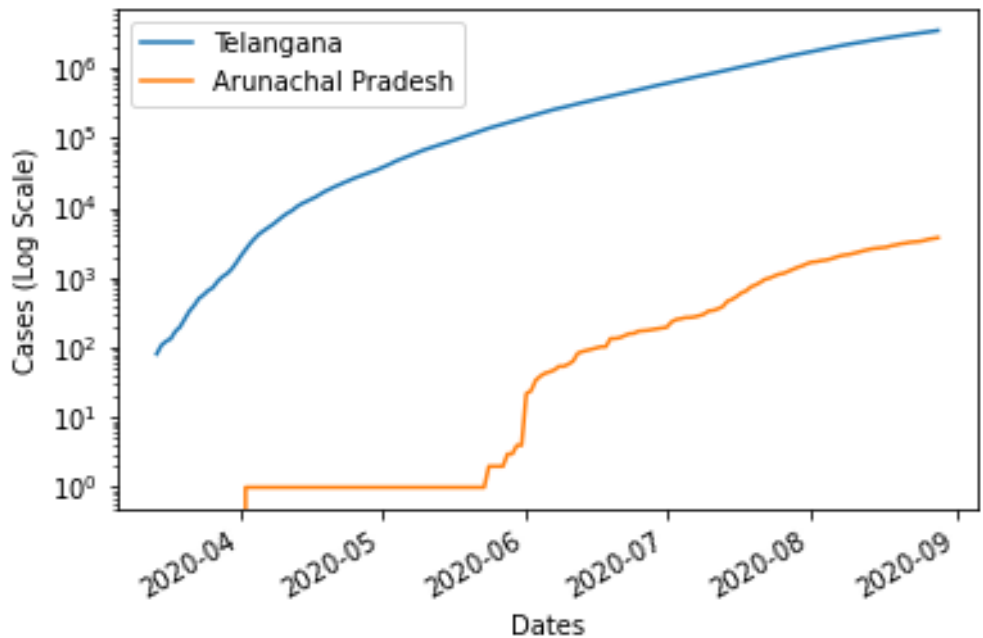

Source: *Covid-19 India Org Data Operations Group*

Graph 7: Confirmed cases in Telangana and Arunachal Pradesh plotted against Days since $P_o$

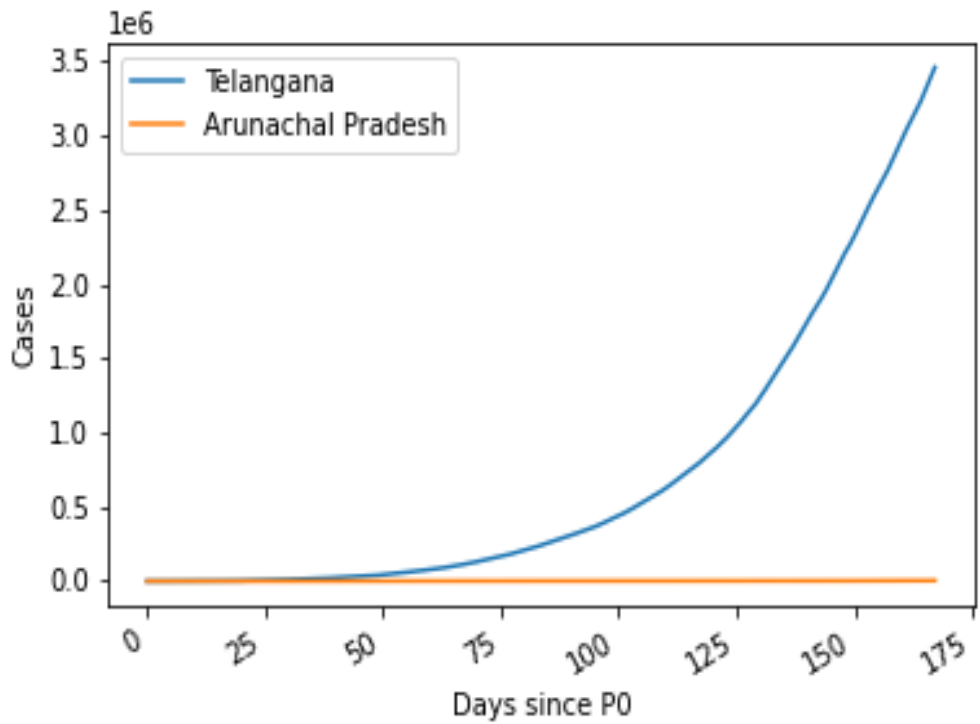

Source: *Covid-19 India Org Data Operations Group*

The two plots, graph 6 and 7, although present the same data, they have minor differences, first, both of these graphs use different X-axis; the first one uses dates however the second one uses 'Days since $P_0$'. This hides the fact that Arunachal Pradesh experienced its first case approximately after 15 days as compared to Telangana, this presents an opportunity to better prepare itself for the surge and thereby handle it much better, the second graph completely shadows this fact and shows an inaccurate view of the states, this leads to formation of incorrect views and opinions

**4.7. Testing is a secret well kept**
It is also to be noted that as the pandemic has stretched over for months. The government's testing efforts have also increased exponentially. This could also hide the fact that the outbreak initially could be bigger than recorded, however due to lack of testing could not be detected in its early stages, this also leads to the infection spreading in the initial stages much more than it was recorded causing the current 'exponential' growth.

Graph 8: Covid – 19 Testing in India

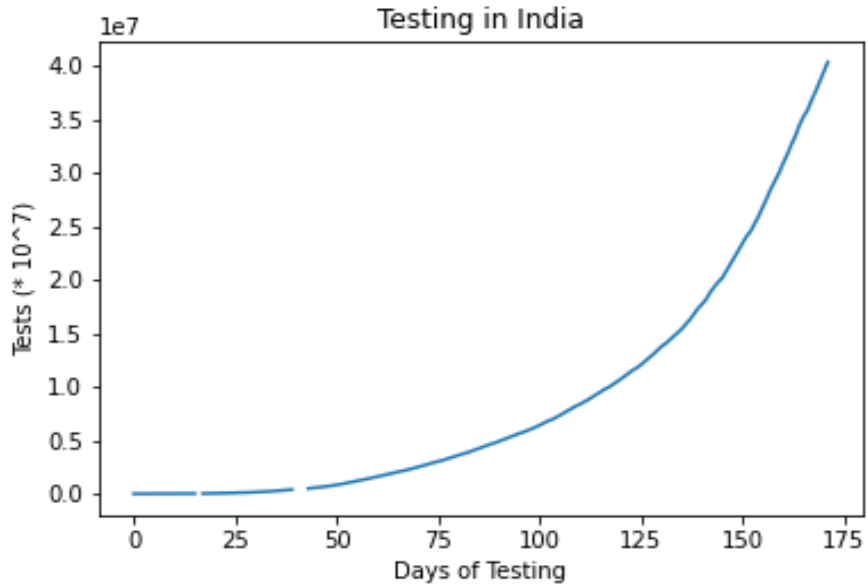

Source: *Covid-19 India Org Data Operations Group*

## 5. Discussion

The mathematical models help in quantifying the probability of infection, while the researcher hopes that the proposed model matches the real data. With respect to that, this epidemic measured using different scales either on X-axis or Y-axis may also change the way in which the data is portrayed. There are two important explanations that can be given to support the use of logarithmic scales while charting or graphing the data. The first explanation that can be given in order to use the logarithmic scale is that it helps to normalize the skewness in the distribution which may have turned out due to few large values in the data. The second point we can raise over here, is that the logarithmic scale helps when one has to represent a percentage change or factors which show multiplicative trends.

In case of usage of graphs for comparing a single phenomenon across two states using same time line may result in wrong depiction and thus should be avoided. In such cases, proper use of variable helps generate accurate output and thus correct interpretation. Through this study, we have shown how the same data could be presented in ways which are unnoticeable to the general public and could therefore deceive them of the complete picture.

**Acknowledgement**
The authors believe in the Open Source Community, all code for the research is publicly available at *github.com/rahulbatra065/covid19-paper*. Please feel free to mail us for any queries regarding the same

**Conflict of Interest**: There is no conflict of interest among the authors

**Funding**: Self-funded

**Ethical approval**: Not applicable

## References

Anastassopoulou, C., Russo, L., Tsakris, A., & Siettosid, C. (2019). *Data-based analysis, modelling and forecasting of the COVID-19 outbreak*, https://doi.org/10.1371/journal.pone.0230405

Babu, M., Marimuthu,  s, Joy, M., Nadaraj, A., Asirvatham, E., & Jeyaseelan, L. (2020). *Forecasting COVID-19 epidemic in India and high incidence states using SIR and logistic growth models*. https://doi.org/10.1016/j.cegh.2020.06.006

Chang, S., Harding, N., Zachreson, C., Cliff, O., & Prokopenko, M. (2020). *Modelling transmission and control of the COVID-19 pandemic in Australia*.

Corman, V. M., Landt, O., Kaiser, M., Molenkamp, R., Meijer, A., Chu, D. K., Bleicker, T., Brünink, S., Schneider, J., Schmidt, M. L., Mulders, D. G., Haagmans, B. L., van der Veer, B., van den Brink, S., Wijsman, L., Goderski, G., Romette, J.-L., Ellis, J., Zambon, M., … Drosten, C. (2020). Detection of 2019 novel coronavirus (2019-nCoV) by real-time RT-PCR. *Euro Surveillance : Bulletin Europeen Sur Les Maladies Transmissibles = European Communicable Disease Bulletin*, *25*(3), 2000045. https://doi.org/10.2807/1560-7917.ES.2020.25.3.2000045

COVID-19 India Org Data Operations Group. (n.d.). *COVID19-India API* . Retrieved August 29, 2020, from https://api.covid19india.org/

Ehrenberg, A. S. C. (1977). Rudiments of Numeracy. *Journal of the Royal Statistical Society. Series A (General)*, *140*(3), 277–297. https://doi.org/10.2307/2344922

Ehrenberg, A. S. C. (1981a). The Problem of Numeracy. *The American Statistician*, *35*(2), 67. https://doi.org/10.2307/2683143

Ehrenberg, A. S. C. (1981b). The Problem of Numeracy. *The American Statistician*, *35*(2), 67–71. https://doi.org/10.1080/00031305.1981.10479310

Fontaine, R. E. (2018). *Describing Epidemiologic Data, Epidemic Intelligence Service, CDC*. https://www.cdc.gov/eis/field-epi-manual/chapters/Describing-Epi-Data.html

Gamero, J., Tamayo, J. A., & Martínez-Román, J. A. (2019). *Forecast of the evolution of the contagious disease caused by novel coronavirus (2019-nCoV) in China*.

Gentleman, J. (1977). Data Reduction: Analysing and Interpreting Statistical Data. *Technometrics*, *139*, 268–269.

Giffen, R., Higgs, H., & Yule, G. U. (1913). *Statistics*. Macmillan and Company, limited. https://books.google.co.in/books?id=6ShBAAAAIAAJ

Goodhardt  1930-, G. J. (Gerald J. (1975). *The television audience : patterns of viewing / [by] G. J. Goodhardt, A. S. C. Ehrenberg, M. A. Collins* (A. S. C. Ehrenberg, M. A. (Martin A. Collins, I. B. Authority, & A. R. Ltd (eds.)). Saxon House ; Lexington Books.

Grimm, R., Boyon, N., & Newall, M. (2016). *How do people across the world trust the news and information they receive from different sources? Trust in the Media*.

Gulankar, A. C. (n.d.). *67% people in India trust media for coronavirus-related news: Survey -*
13

# Appendix

**Data Sheets**

Table 1: Covid-19 related indicators for India

|  | Daily Confirmed | Total Confirmed | Daily Recovered | Total Recovered | Daily Deceased | Total Deceased |
|---|---|---|---|---|---|---|
| **Total Observations (Count)** | 197 | 197 | 197 | 197 | 197 | 197 |
| **Mean (Average)** | 12485.41 | 378625.51 | 8886.44 | 236708.47 | 244.44 | 9181.57 |
| **Standard Deviation** | 18176.75 | 607071.56 | 14381.19 | 410026.07 | 315.21 | 13161.70 |
| **Minimum Value Observed** | 0 | 1 | 0 | 0 | 0 | 0 |
| **25th Percentile** | 27 | 198 | 2 | 20 | 1 | 4 |
| **50th Percentile** | 3344 | 56351 | 1295 | 16776 | 103 | 1889 |
| **75th Percentile** | 18205 | 491193 | 12064 | 285672 | 414 | 15309 |
| **Maximum Value Observed** | 67066 | 2459626 | 57759 | 1750629 | 2004 | 48156 |

Source: *Covid-19 India Org Data Operations Group*

Table 2: State-wise Covid-19 related indicators (Confirmed Cases)

|  | Maharashtra | Kerala | Karnataka |
|---|---|---|---|
| **Total Observations (Count)** | 168 | 168 | 168 |
| **Mean (Average)** | 4452.35 | 412.52 | 1897.33 |
| **Standard Deviation** | 4350.37 | 637.24 | 2739.65 |
| **Minimum Value Observed** | 3 | 0 | 0 |
| **25th Percentile** | 552 | 10.75 | 17 |
| **50th Percentile** | 2762.5 | 80.5 | 214.5 |
| **75th Percentile** | 8016 | 642.75 | 3660 |
| **Maximum Value Observed** | 14888 | 2543 | 9386 |

Source: *Covid-19 India Org Data Operations Group*